\documentclass[12pt, draftclsnofoot, onecolumn]{IEEEtran}
\usepackage[T1]{fontenc}
\usepackage{cite}
\ifCLASSINFOpdf
\else
\fi   
\usepackage{url}
\usepackage{amsbsy}
\usepackage{amsmath}
\usepackage{fixmath}
\usepackage{amssymb}
\usepackage{ulem}

\usepackage[noend]{algorithmic}
\usepackage{algorithm}
\newlength\myindent
\algsetup{indent=\myindent}
\algsetup{linenodelimiter=}

\usepackage{tikz}
\usetikzlibrary{arrows,decorations,backgrounds,shadows,plotmarks}

\usepackage{amsmath}
\usepackage{adjustbox}
\usepackage{romannum}
\usepackage{booktabs}
\usepackage{caption}
\usepackage{lipsum}
\usepackage{xcolor,colortbl}
\usepackage{makecell}
\usepackage{pgfplots}
\usepackage{bm}

\newtheorem{assumption}{Assumption}
\usepackage{algorithm}
\usepackage{algorithmic}
\usepackage{siunitx}
\usepackage{float}
\usepackage{pgfplots}
\pgfplotsset{compat=newest}
\pgfplotsset{plot coordinates/math parser=false}
\pgfplotsset{every axis/.append style={font=\footnotesize}}
\newlength\figureheight
\newlength\figurewidth
\definecolor{TUMBeamerYellow}    {rgb} {1.000,0.706,0.000}    
\definecolor{TUMBeamerOrange}    {rgb} {1.000,0.502,0.000}    
\definecolor{TUMBeamerRed}       {rgb} {0.898,0.204,0.094}    
\definecolor{TUMBeamerDarkRed}   {rgb} {0.792,0.129,0.247}    
\definecolor{TUMBeamerBlue}      {rgb} {0.000,0.600,1.000}    
\definecolor{TUMBeamerLightBlue} {rgb} {0.255,0.745,1.000}    
\definecolor{TUMBeamerGreen}     {rgb} {0.569,0.675,0.420}    
\definecolor{TUMBeamerLightGreen}{rgb} {0.710,0.792,0.510}    

\usepackage{gensymb}
\usepackage{amsmath,graphicx}
\usepackage{tikz}
\usepackage{amssymb}
\usepackage{trfsigns}
\usepackage{siunitx}
\usepackage{bm}
\usepackage{pgfplots}
\usepackage{float}
\usepackage{algorithm}
\usepackage{algorithmic}
\pgfplotsset{compat=newest}
\pgfplotsset{plot coordinates/math parser=false}
\pgfplotsset{every axis/.append style={font=\footnotesize}}

\usepackage{cite}
\usepackage{amsmath,amssymb,amsfonts}
\usepackage{algorithmic}
\usepackage{graphicx}
\usepackage{textcomp}
\usepackage{xcolor}
\def\BibTeX{{\rm B\kern-.05em{\sc i\kern-.025em b}\kern-.08em
		T\kern-.1667em\lower.7ex\hbox{E}\kern-.125emX}}

\begin{document}
\title{Reproducible Evaluation of Neural Network Based Channel Estimators And Predictors Using A Generic Dataset}
\author{Nurettin~Turan and Wolfgang~Utschick\thanks{The authors are with Methods of Signal Processing, Technische Universit\"at M\"unchen, 80290 Munich, Germany (e-mail: \{nurettin.turan,utschick\}@tum.de.}
}
\markboth{}{}
	
\maketitle

\begin{abstract}
A low-complexity neural network based approach for channel estimation was proposed recently, where assumptions on the channel model were incorporated into the design procedure of the estimator. Instead of using data from a measurement campaign as done in previous work, we evaluate the performance of the convolutional neural network (CNN) based channel estimator by using a reproducible mmWave environment of the DeepMIMO dataset. We further propose a neural network based predictor which is derived by starting from the linear minimum mean squared error (LMMSE) predictor. We start by deriving a weighted sum of LMMSE predictors which is motivated by the structure of the optimal MMSE predictor. This predictor provides an initialization (weight matrices, biases and activation function) to a feed-forward neural network based predictor. With a properly learned neural network, we show that it is possible to easily outperform the LMMSE predictor based on the Jakes assumption of the underlying Doppler spectrum in an reproducible indoor scenario of the DeepMIMO dataset.
\end{abstract}

\begin{IEEEkeywords}
channel state information, minimum mean squared error estimation / prediction, machine learning, neural networks
\end{IEEEkeywords}

\section{Introduction}
\label{sec:intro}

In \cite{Neumann} a low-complexity neural network based approach for channel estimation was derived by incorporating channel model assumptions into the design procedure of the estimator \cite{Hellings}. The authors of \cite{Neumann} derived the neural network based estimator by starting from the 3GPP channel model and showed further, that the neural network based channel estimator is optimal for the 3GPP model with one propagation path. For more general 3GPP system setups with more than one propagation path, the powerful neural networks with optimization parameters allowed the construction of channel estimators by using machine learning methods to train these optimization variables \cite{Hellings}. In \cite{Hellings} the performance of the neural network based channel estimator was verified with data from a measurement campaign. In the following, we want to verify the performance of the neural network based estimator by using the DeepMIMO dataset in a mmWave environment \cite{Alkhateeb2019}. 

For increasing the achievable transmission rate in a wireless communication system, it is beneficial to have channel state information (CSI) at the transmitter side \cite{Zemen}. In scenarios, where the users are moving, the CSI may get outdated rapidly; thus, channel prediction plays an important role \cite{Zemen}. After reformulating the general expression of the linear MMSE (LMMSE) predictor, a similar approach as for the learning based low complexity MMSE channel estimator \cite{Neumann} is used to obtain a learning based MMSE channel predictor \cite{turan2019learning}. The starting point to the channel prediction problem is an underlying channel model. By reformulating the LMMSE predictor and by further making two key assumptions it is possible to derive a weighted sum of LMMSE predictors which has the structure of a feed-forward neural network \cite{turan2019learning}. After training this feed-forward neural network, it is possible to easily outperform the LMMSE predictor based on the Jakes assumption of the underlying Doppler spectrum in an indoor scenario of the DeepMIMO dataset \cite{turan2019learning}.

By using the DeepMIMO dataset, we ensure the reproducibility of our work, which is for machine learning applications in the field of communications often not the case. 
\section{Channel Estimation based on Convolutional Neural Networks}

We consider an uplink scenario, where a single antenna user transmits data to a base station (BS) with $M$ antenna elements. The noisy observation at the BS are given by\footnote{We assume that the coherence time is $T=1$.} \cite{Neumann, Hellings, Koller}: $\mathbf{y} = \mathbf{h} + \mathbf{n}$ where the complex additive white Gaussian noise (AWGN) is described by:
$\mathbf{n} \sim \mathcal{N}_\mathbb{C}(\mathbf{0},\mathbf{\Sigma}_\mathbf{n} = \sigma_n^2 \mathbf{I}_M).$ Given a set of parameters $\boldsymbol{\delta} \sim p(\boldsymbol{\delta})$ the channel $\mathbf{h}$ is assumed to be conditionally Gaussian distributed \cite{Neumann,Hellings,Koller}: $\mathbf{h}|\boldsymbol{\delta} \sim \mathcal{N}_\mathbb{C}( \mathbf{0},\mathbf{\Sigma}_{\boldsymbol{\delta}}).
$ In the 3GPP urban (micro or macro) scenarios, the receive signal is constructed by the superposition of a few propagation clusters, where each cluster includes many subpaths \cite{Neumann,Hellings,Koller}. Each of the clusters has an angle center and a specific gain, which are collected in $\boldsymbol{\delta}$ \cite{Neumann,Hellings,Koller}. For the case that the parameter ${\boldsymbol{\delta}}$ is known, the linear minimum mean squared error (LMMSE) channel estimate can be calculated with \cite{Neumann,Hellings,Koller}:
\begin{equation}
    \hat{\mathbf{h}} = \mathbf{W}_{\boldsymbol{\delta}} \mathbf{y}, \ \text{with} \ \ 
    \mathbf{W}_{\boldsymbol{\delta}} = \mathbf{\Sigma}_{\boldsymbol{\delta}} (\mathbf{\Sigma}_{\boldsymbol{\delta}} + \sigma_n^2 \mathbf{I}_M )^{-1}.
    \label{LCEsampleFilter}
\end{equation}
Since the parameter ${\boldsymbol{\delta}}$ is unknown and described by a prior distribution $p({\boldsymbol{\delta}})$, an estimate of the channel vector can be obtained by\cite{Neumann,Hellings,Koller}:
\begin{align}
        \hat{\mathbf{h}} 
        &= E[\mathbf{h}|\mathbf{y}] = E[E[\mathbf{h}|\mathbf{y},{\boldsymbol{\delta}}]|\mathbf{y}] = E[\mathbf{W}_{\boldsymbol{\delta}} \mathbf{y} | \mathbf{y}]\\
        &= E[\mathbf{W}_{\boldsymbol{\delta}} | \mathbf{y}] \mathbf{y} = \hat{\mathbf{W}}_{\text{MMSE}}(\mathbf{y}) \mathbf{y},
\end{align}
where the filter $\hat{\mathbf{W}}_{\text{MMSE}}(\mathbf{y})$ is given by \cite{Neumann,Hellings,Koller}:
\begin{align}
    \hat{\mathbf{W}}_{\text{MMSE}}(\mathbf{y}) 
    &= E[\mathbf{W}_{\boldsymbol{\delta}} | \mathbf{y}] = \dfrac{\int p(\mathbf{y|{\boldsymbol{\delta}}}) \mathbf{W}_{\boldsymbol{\delta}} p(\boldsymbol{\delta}) d{\boldsymbol{\delta}}}{\int p(\mathbf{y|{\boldsymbol{\delta}}}) p(\boldsymbol{\delta}) d{\boldsymbol{\delta}}}.
\end{align}
By defining $\hat{\mathbf{C}} = {\sigma^{-2}_n} \mathbf{y}\mathbf{y}^{H}$, we can express the estimated filter $\hat{\mathbf{W}}_{\text{MMSE}}$ now depending  on $\hat{\mathbf{C}}$ with \cite{Neumann}:
\begin{equation}
\hat{\mathbf{W}}_\text{MMSE}(\hat{\mathbf{C}}) =  \dfrac{\int \exp{(\mathrm{tr}( {\mathbf{W}}_{\boldsymbol{\delta}} \hat{\mathbf{C}})}+b_{\boldsymbol{\delta}}) {\mathbf{W}}_{\boldsymbol{\delta}} p({\boldsymbol{\delta}})  d{\boldsymbol{\delta}} } { \int \exp{(\mathrm{tr}( {\mathbf{W}}_{\boldsymbol{\delta}} \hat{\mathbf{C}})}+b_{\boldsymbol{\delta}}) p( {\boldsymbol{\delta}}) d{\boldsymbol{\delta}} }.
\end{equation}
The above expression is not computationally tractable. Thus, the authors of \cite{Neumann} made several assumptions: First, they assumed that the prior is discrete and uniform. Then, a uniform linear array (ULA) with a relatively high number of antennas was assumed, which allowed to approximate all possible covariance matrices and their corresponding filters by \cite{Neumann, Hellings}:
\begin{equation}
    {\mathbf{C}}_{\boldsymbol{\delta}} = \mathbf{Q}^H \mathrm{diag}(\mathbf{c}_{\boldsymbol{\delta}}) \mathbf{Q} \  \ \ \ {\mathbf{W}}_{\boldsymbol{\delta}} = \mathbf{Q}^H \mathrm{diag}(\mathbf{w}_{\boldsymbol{\delta}}) \mathbf{Q},
\end{equation}
Possible candidates for $\mathbf{Q}$ are either the Discrete Fourier transform (DFT) matrix, $\mathbf{Q} = \mathbf{F}_1 \in  \mathbb{C}^{M \times M}$ (Circulant approx.) or the first M columns of the $2M\times 2M$ DFT matrix, $\mathbf{Q} = \mathbf{F}_2 \in  \mathbb{C}^{2M \times M}$ (Toeplitz approx.) \cite{Neumann, Gray2006}. By further considering only one cluster, the spectrum of the channel becomes shift invariant \cite{Neumann}. With these assumptions in hand, the channel estimator is \cite{Neumann}:
\begin{align}
    \hat{\mathbf{W}}(\hat{\mathbf{c}}) &= \mathbf{Q}^H \mathrm{diag}(\hat{\mathbf{w}}(\hat{\mathbf{c}})) \mathbf{Q},  \label{LCEFastEstimator} \\
    \text{with} \ \  \hat{\mathbf{w}}(\hat{\mathbf{c}}) &= \mathbf{w}_0 * \text{softmax}(\tilde{\mathbf{w}}_0*\hat{\mathbf{c}}+\mathbf{b}),
\end{align}
where ${\mathbf{w}}_0$ and ${\mathbf{b}}$ are parameters that depend on the set of possible covariance matrices $ {\mathbf{C}}_{\boldsymbol{\delta}}$ and $\tilde{\mathbf{w}}_0$ is ${\mathbf{w}}_0$ in reversed order \cite{Neumann,Hellings}. $\hat{\mathbf{c}} = {\sigma^{-2}_n} |\mathbf{Qy}|^2$, and the softmax function is defined as $\text{softmax}(\mathbf{x}) = \frac{ \exp{( \mathbf{x} )} } { \mathbf{1}^T\exp{(\mathbf{x})}}$. The estimator from \eqref{LCEFastEstimator} does not yield the MMSE estimator in most practical scenarios, since the three key assumptions made above are approximations, which do only hold for specific cases. However, the estimator from \eqref{LCEFastEstimator} has a complexity of $\mathcal{O}(M \text{log}M)$ and has the structure of a convolutional neural network (CNN) with one hidden layer and the softmax activation function \cite{Neumann}. The idea is now to use a CNN in order to have weights and biases, which are learned by training, to compensate the approximation errors made with the three key assumptions \cite{Neumann}. Thus, the CNN based estimator is \cite{Neumann}:
\begin{align}
    \hat{\mathbf{W}}_\text{CNN}(\hat{\mathbf{c}}) &= \mathbf{Q}^H \mathrm{diag}(\hat{\mathbf{w}}_{\text{CNN}}(\hat{\mathbf{c}})) \mathbf{Q}, \\
    \text{with} \ \  \hat{\mathbf{w}}_{\text{CNN}}(\hat{\mathbf{c}}) &= \mathbf{a}_1 * \text{softmax}({\mathbf{a}}_2*\hat{\mathbf{c}}+\mathbf{b}_1) + \mathbf{b}_2,
\end{align}
where $\mathbf{a}_1$, $\mathbf{a}_2$, $\mathbf{b}_1$ replace the variables of the estimator from \eqref{LCEFastEstimator} and additionally a second bias term $\mathbf{b}_2$ is included. These variables are optimized via stochastic gradient descent with the Mean Squared Error (MSE) as cost function \cite{Neumann,Hellings,Koller}:
\begin{equation}
    \min_{\mathbf{a}_1, \mathbf{a}_2, \mathbf{b}_1, \mathbf{b}_2 } E[\| \mathbf{h}- \hat{\mathbf{W}}_\text{CNN}(\hat{\mathbf{c}}) \mathbf{y} \|_F^2]
\end{equation}

\subsection{CNN based Channel Estimation for a mmWave System}

In this section, we use the CNN based channel estimator from above in a mmWave environment by using the DeepMIMO dataset. The accurate channel vectors provided by the DeepMIMO dataset are constructed by using a ray-tracing tool \cite{Remcom}. The channel vectors are normalized such that $E[\| \mathbf{h} \|^2] = M$. Given the accurate channel vectors, we obtain noisy versions by adding AWGN as described above. Accordingly, the signal-to-noise ratio (SNR) is defined as: $\text{SNR} = 10\ \text{log}_{10}({ \sigma_n^{-2}})$. The normalized MSE (NMSE) is: $\text{NMSE} = \frac{E[\| \mathbf{h} - \hat{\mathbf{h}} \|^2]}{M}$, where $\hat{\mathbf{h}}$ is an estimate of the true channel ${\mathbf{h}}$. In the following, we use the ray-tracing scenario "O1" of the DeepMIMO dataset with a carrier frequency of $60\si{GHz}$. We only consider one carrier and a limited section of the User Grid 1. The first active row is R700 and the last active row is R1418. In this limited section the horizontal and vertical distances between neighbouring grid points are $20\si{cm}$. The obtained channel vectors of each grid point are randomly assigned either to a training set or a test set. Thus, we have 6000 training batches, where each consists of 20 channel realizations and 100 test batches, with each including 100 channel vectors. We applied the same hierarchical learning strategy as presented in \cite{Neumann} and shuffle the train dataset prior to each hierarchical learning step.

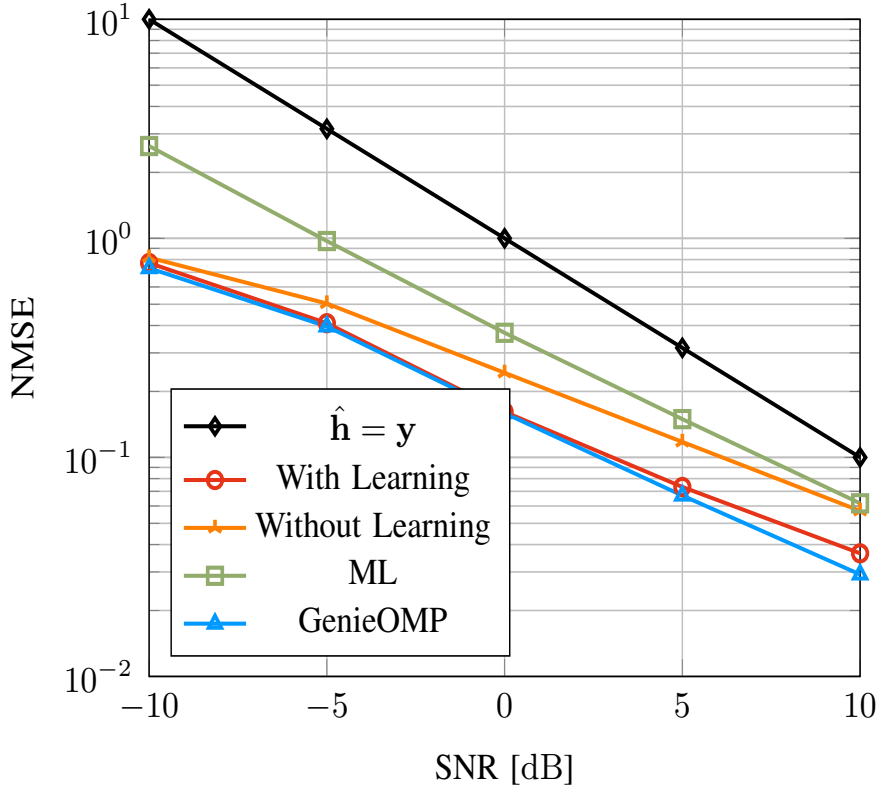
\begin{figure}[t]
    \centering
    \resizebox{350pt}{308pt}{
    \begin{tikzpicture}
\begin{axis}[
legend pos= south west,
legend style={font=\footnotesize},
xmin=-10,
xmax=10,
ymin=1e-2,
ymax=1e+1,
xlabel={SNR [\si{dB}]},
ylabel={NMSE},
grid=both,
ymode=log
]
\addplot[line width=1,solid,mark=diamond]
table[x=snr,y=nmse] {LCE_LOS_EstEqualObs.txt};
\addlegendentry{$\hat{\mathbf{h}} = \mathbf{y}$};
\addplot[line width=1,TUMBeamerRed,solid,mark=o]
table[x=snr,y=nmse] {LCE_LOS_WithLearning.txt};
\addlegendentry{With Learning};
\addplot[line width=1,TUMBeamerOrange,solid,mark=Mercedes star]
table[x=snr,y=nmse] {LCE_LOS_WithoutLearning.txt};
\addlegendentry{Without Learning};
\addplot[line width=1,TUMBeamerGreen,solid,mark=square]
table[x=snr,y=nmse] {LCE_LOS_ML.txt};
\addlegendentry{ML};
\addplot[line width=1,TUMBeamerBlue,solid,mark=triangle]
table[x=snr,y=nmse] {LCE_LOS_GenieOMP.txt};
\addlegendentry{GenieOMP};
\end{axis}
\end{tikzpicture}
}
\caption{NMSE using a 16-antenna ULA at BS3.}
\label{fig:LOS}
\end{figure}

In the first simulation (Fig. \ref{fig:LOS}) we have a 16-antenna ULA $(M=16)$ with an antenna spacing of $\lambda/2$ at BS3 and the maximum number of paths is set to 25 (all available paths are used to construct the channel vector of each grid point). 
As benchmark we plot channel estimators based on an approximate maximum likelihood (ML) estimation of the covariance matrix, which are constructed by making similar assumptions on the channels as above \cite{Neumann, Hellings, Neumann2}, and the compressive sensing method orthogonal matching pursuit (OMP) \cite{OMP}, which depends on choice of the sparsity level. This choice is performed with the knowledge of the true channel, thus we refer to this as \textit{GenieOMP}, because of the optimistic performance evaluation \cite{Neumann, Hellings}. For further readings on these benchmark algorithms, the reader is referred to \cite{Neumann}. By learning the CNN channel estimator, an improvement as compared to the estimator without learning  \eqref{LCEFastEstimator} can be achieved (Fig. \ref{fig:LOS}). The CNN based estimator achieves a similar performance as the GenieOMP estimator and outperforms the ML based channel estimator. 

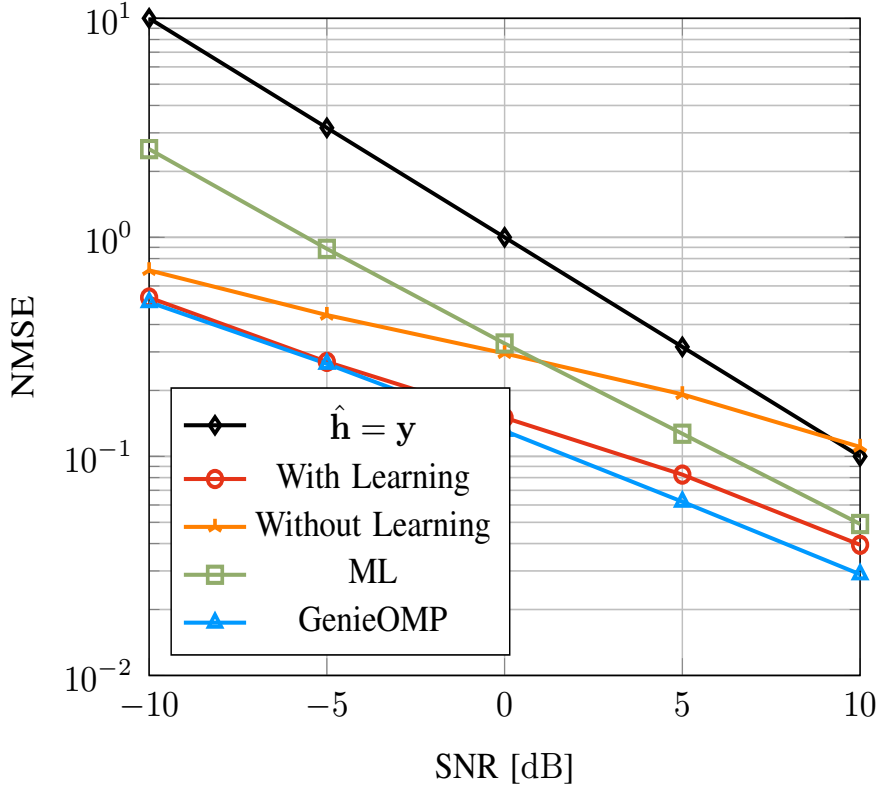
\begin{figure}[t]
    \centering
    \resizebox{350pt}{308pt}{
    \begin{tikzpicture}
\begin{axis}[
legend pos= south west,
legend style={font=\footnotesize},
xmin=-10,
xmax=10,
ymin=1e-2,
ymax=1e+1,
xlabel={SNR [\si{dB}]},
ylabel={NMSE},
grid=both,
ymode=log
]
\addplot[line width=1,solid,mark=diamond]
table[x=snr,y=nmse] {LCE_NLOS_EstEqualObs.txt};
\addlegendentry{$\hat{\mathbf{h}} = \mathbf{y}$};
\addplot[line width=1,TUMBeamerRed,solid,mark=o]
table[x=snr,y=nmse] {LCE_NLOS_WithLearning.txt};
\addlegendentry{With Learning};
\addplot[line width=1,TUMBeamerOrange,solid,mark=Mercedes star]
table[x=snr,y=nmse] {LCE_NLOS_WithoutLearning.txt};
\addlegendentry{Without Learning};
\addplot[line width=1,TUMBeamerGreen,solid,mark=square]
table[x=snr,y=nmse] {LCE_NLOS_ML.txt};
\addlegendentry{ML};
\addplot[line width=1,TUMBeamerBlue,solid,mark=triangle]
table[x=snr,y=nmse] {LCE_NLOS_GenieOMP.txt};
\addlegendentry{GenieOMP};
\end{axis}
\end{tikzpicture}
}
\caption{NMSE using a 16-antenna ULA at BS17.}
\label{fig:NLOS}
\end{figure}

In the next simulation (Fig. \ref{fig:NLOS}) we use the same simulation setting and street section, but this time at BS17. In this way, we have no line of sight components as compared to the previous simulation, where BS3 is placed at the considered street section \cite{Alkhateeb2019}.
We can see that the assumptions to construct the channel estimator from \eqref{LCEFastEstimator} are not as well suited as for the setting from before, because of the performance curve of the untrained channel estimator from \eqref{LCEFastEstimator}, especially for large SNR values. However, we can see that the trained CNN adapts to the environment and again easily outperforms the ML based estimator and achieves a similar performance as the GenieOMP estimator.

\section{Channel Prediction based on feed-forward neural networks}
\label{sec:chmod}

In this section we derive a learning based channel predictor and evaluate its performance in an indoor scenario of the DeepMIMO dataset. We start with a general channel model, which is constructed by the superposition of $P$ impinging plane-waves at a moving user with constant velocity $v$ ~\cite{Zemen,Clarke}. Each of these paths is mainly determined by a path-specific Doppler shift $f_p$ and path phase $\psi_p$, where the Doppler shift of path $p$ equals to: $f_p = \cos{(\delta_p)} {f_c v}/{c}$, with $f_c$ being the carrier frequency, $c$ the speed of light and $\delta_p$ the direction of arrival (DoA) of path $p$. The maximum possible Doppler shift is defined as the Doppler bandwidth $B_D = {v f_c}/ {c}$. We assume that over a block of $M+N$ symbols the path phases and the Doppler shifts remain unchanged, where $M$ is the \textit{observation length}, $N$ is the \textit{prediction length} and $T_s$ is the \textit{symbol duration}, which is much longer than the delay spread of the channel, thus, we have a frequency-flat channel~\cite{Zemen}. We further assume that each path-phase $\psi_p$ and each path specific DoA $\delta_p$ are uniformly distributed \cite{Zemen, Clarke}. Thus, following the argumentation in \cite{Zemen} the channel coefficients $h[m]$ for $m=0,\dots, M+N-1$ are constructed by:
\begin{equation}
    h[m] = \sum_{p=0}^{P-1} ({1}/{\sqrt{P}}) e^{j\psi_p} e^{j2\pi f_p T_s m} = \sum_{p=0}^{P-1} a_p e^{j2\pi f_p T_s m}.
    \label{channelcoeff}
\end{equation}
Drawing the limit of $P\to\infty$ the channel coefficiencts follow a Gaussian distribution based on the central limit theorem. However, it is important to note that for a low number of paths the obtained channel coefficients are distributed non-Gaussian.

The goal is to make use of the correlations between the channel coefficients, in order to predict any desired channel coefficient in the \textit{prediction interval} $\mathcal{I}_N = \{M,M+1,\dots, M+N-1\}$, from noisy observations of the channel coefficients of the \textit{observation interval} $\mathcal{I}_M = \{0,1,\dots, M-1\}$. The channel model is a time-variant block-fading model. The zero mean and unit variance process is wide-sense stationary over a block, which consists of the union of the observations interval $\mathcal{I}_M$ and the prediction interval $\mathcal{I}_N$ \cite{Zemen, Zemen3}. The power spectral density (PSD) is given by the weighted sum of Dirac pulses at the Doppler shift frequencies $f_p$~\cite{Zemen, Goldsmith}, i.e., $ S_h(f) = \sum_{p=0}^{P-1} |a_p|^2 \delta(f-f_p) $. Consequently, the discrete covariance function $R_h[k]$ is obtained by sampling the inverse Fourier transform of the PSD \cite{Goldsmith, Bracewell2000}, viz., $
R_h[k] = \sum_{p=0}^{P-1} |a_p|^2 e^{j 2 \pi f_p T_s k} $ at $k = 0, 1, \dots, M+N-1$. With the specific assumptions of the channel model in \eqref{channelcoeff} and if there are infinitely many paths, i.e., $P\to\infty$, the limit of the discrete covariance functions $R_h[k]$ is equal to $ J_0(2\pi k T_s f_c {v}/{c})$~\cite{Zemen,Goldsmith,Jakes}. By collecting the channel coefficients $h[m]$ of the observation interval $\mathcal{I}_M$ in a vector $\mathbf{h} = [h[M-1], h[M-2], \dots, h[1], h[0]]^{T}$, we obtain the covariance matrix $\mathbf{\Sigma}_{\mathbf{h}}$  ~\cite{Zemen}:
\begin{equation}
    \footnotesize{
    \mathbf{\Sigma}_{\mathbf{h}} = 
\begin{bmatrix}
R_h[0]& R_h[1]      & \dots & R_h[M-1] \\  
R_h^*[1] & R_h[0]     & \dots & R_h[M-2] \\
\vdots   & \vdots & \ddots & \vdots \\
R_h^*[M-2] & R^*_h[M-3]  &   \dots & R_h[1] \\
R_h^*[M-1] & R^*_h[M-2]  &   \dots & R_h[0] \\
\end{bmatrix}}.
\label{covmatstandard}
\end{equation}

\subsection{LMMSE Predictor}
\label{sec:pagestyle}

The noisy observations of channel coefficients within the observation interval $\mathcal{I}_M$ are collected in a vector $\mathbf{y} = \mathbf{h} + \mathbf{n}$, where the complex additive white Gaussian noise (AWGN) is described by $\mathbf{n} \sim \mathcal{N}_\mathbb{C}(\mathbf{0},\mathbf{\Sigma}_\mathbf{n} = \sigma_n^2 \mathbf{I}_M)$, with $\mathbf{I}_M$ being the $M\times M$ identity matrix. Accordingly, the covariance matrix of the noisy observations $\mathbf{y}$ is $\mathbf{\Sigma}_\mathbf{y} = \mathbf{\Sigma}_{\mathbf{h}}  + \sigma_n^2 \mathbf{I}_M$. With these quantities in hand, channel coefficients of the prediction interval $\mathcal{I}_N$ can be obtained with the $l-$step LMMSE predictor~\cite{Zemen,Kay}:
\begin{equation}
\hat{h}[m] = \hat{h}_m = \mathbf{c}_{h_{m}\mathbf{y}}^{H} \mathbf{\Sigma}_{\mathbf{y}}^{-1} \mathbf{y},
\label{lstepWP}
\end{equation}
with $m \in \mathcal{I}_N$ and $l = m-(M-1)$ and the correlation vector $ \mathbf{c}_{h_{m}\mathbf{y}}^{H}$ equal to
\begin{align}
 \mathbf{c}_{h_{m}\mathbf{y}}^{H} = [R_h[l], R_h[1+l], \dots, R_h[M-1+l]].
\end{align}

We derive a reformulated version of the LMMSE predictor in the following. For a fixed step length $l$ the vector of channel coefficients $\mathbf{h}$ of the observation interval $\mathcal{I}_M$ is artificially extended by $l$ channel coefficients of the prediction interval $\mathcal{I}_N$: $\mathbf{h}^{l-\text{ext}} = [{h[m]},{h[m-1]}\dots,{h[M]} ,\mathbf{h}^T]^T,$ with $m \in \mathcal{I}_N$ and $m = l+(M-1)$. The extended covariance matrix $\mathbf{\Sigma}_\mathbf{h}^{l-\text{ext}}$ is constructed analogous to the covariance matrix $\mathbf{\Sigma}_{\mathbf{h}}$ from \eqref{covmatstandard}.
By construction, the correlation vector $\mathbf{c}_{h_{m}\mathbf{y}}^{H}$ is identical to the zeroth row starting from the $l-$th column of the extended covariance matrix $\mathbf{\Sigma}_\mathbf{h}^{l-\text{ext}}$ and the covariance matrix $\mathbf{\Sigma}_{\mathbf{h}}$ is embedded in the bottom right part of the extended covariance matrix $\mathbf{\Sigma}_\mathbf{h}^{l-\text{ext}}$. With the following definitions:
\begin{equation}
       \mathbf{e}_1^T = [1,0, \dots, 0] \hspace{1.5cm} (1 \times M+l)
\end{equation}
\begin{equation}
       \mathbf{S} = 
       \begin{bmatrix}
        \mathbf{0} \\
        \mathbf{I}_M
\end{bmatrix} \hspace{1.5cm} (M+l \times M).
\end{equation}
the correlation vector $\mathbf{c}_{h_{m}\mathbf{y}}^{H}$ and the covariance matrix can be extracted $\mathbf{\Sigma}_{\mathbf{h}}$ from the extended covariance matrix $\mathbf{\Sigma}_\mathbf{h}^{l-\text{ext}}$ by:
\begin{equation}
       { \mathbf{c}_{h_{m}\mathbf{y}}^{H} } = \mathbf{e}_1^T \mathbf{\Sigma}_\mathbf{h}^{l-\text{ext}} \mathbf{S}
       \label{corrvecreformu} \quad \text{and} \quad 
       {\mathbf{\Sigma}_{\mathbf{h}}} = \mathbf{S}^T \mathbf{\Sigma}_\mathbf{h}^{l-\text{ext}} \mathbf{S}.
\end{equation}
We are now able to reformulate the $l-$step LMMSE predictor from \eqref{lstepWP} by incorporating our results from \eqref{corrvecreformu} \cite{turan2019learning}:
\begin{align}
    \hat{h}_m 
      & = \mathbf{e}_1^T \mathbf{\Sigma}_\mathbf{h}^{l-\text{ext}} \mathbf{S} (\mathbf{S}^T \mathbf{\Sigma}_\mathbf{h}^{l-\text{ext}} \mathbf{S} +  \sigma_n^2 \mathbf{I}_M)^{-1} \mathbf{y}
      \label{lstepWPreformu} \\
      & = \mathbf{e}_1^T\mathbf{W}^{l-\text{ext}}\mathbf{y}. 
      \label{lstepWPreformu2}
\end{align}

\subsection{Gridded Predictor}
\label{sec:typestyle}
In the following, we use Bayes' approach of \cite{Neumann} to derive an approximated MMSE predictor. The proposed solution is based on the random variables ${\boldsymbol{\delta}}$ described by a given distribution $p({\boldsymbol{\delta}})$ and the assumption that for each sample of ${\boldsymbol{\delta}}$ the closed-form solution $ {\mathbf{W}}_{\boldsymbol{\delta}} $ of the LMMSE predictor according to $ \mathbf{W}^{l-\text{ext}} $ as in \eqref{lstepWPreformu2} is available \cite{Neumann, Koller, Hellings}:
\begin{equation}
       \hat{\mathbf{W}}_\text{MMSE} = \int p({\boldsymbol{\delta}}|\mathbf{y}) {\mathbf{W}}_{\boldsymbol{\delta}} d{\boldsymbol{\delta}}.
\end{equation}
Note that each realization of the random ${\boldsymbol{\delta}}$ corresponds to the DoAs of a sampled scenario that determines the path-specific Doppler shift.
By using Bayes' theorem, it follows for the estimated filter:
\begin{equation}
       \hat{\mathbf{W}}_\text{MMSE} = \dfrac{\int p(\mathbf{y}|{\boldsymbol{\delta}}) {\mathbf{W}}_{\boldsymbol{\delta}} p({\boldsymbol{\delta}})  d{\boldsymbol{\delta}} } { \int p(\mathbf{y}|{\boldsymbol{\delta}}) p( {\boldsymbol{\delta}}) d{\boldsymbol{\delta}} }.
       \label{estFilt}
\end{equation}
The likelihood of $\mathbf{y}$ given ${\boldsymbol{\delta}}$ is assumed to be Gaussian:\footnote{We now indexed the second order statistical moments with $\boldsymbol{\delta}$ to express the dependency on the selected sample of the scenario.}
\begin{equation}
  p(\mathbf{y}|{\boldsymbol{\delta}}) \propto {\exp{(-\mathbf{y}^H\mathbf{\Sigma_{y_{\boldsymbol{\delta}}}^{-1}}\mathbf{y})}}/{|\mathbf{\Sigma_{y_{\boldsymbol{\delta}}}}|}.
\end{equation}
We now wish to express $\mathbf{\Sigma_{y_{\boldsymbol{\delta}}}^{-1}}$ in terms of ${\mathbf{W}}_{\boldsymbol{\delta}}$. To this end, we firstly identify ${\mathbf{W}}_{\boldsymbol{\delta}}$ with the predictor in \eqref{lstepWPreformu} \cite{turan2019learning}:
\begin{equation}
       \mathbf{\Sigma}_\mathbf{h_{\boldsymbol{\delta}}}^{l-\text{ext}} \mathbf{S} (\mathbf{S}^T \mathbf{\Sigma}_\mathbf{h_{\boldsymbol{\delta}}}^{l-\text{ext}} \mathbf{S} + \sigma_n^2\mathbf{I}_M)^{-1} = {\mathbf{W}}_{\boldsymbol{\delta}} 
\end{equation}
\begin{equation}
       \mathbf{\Sigma}_\mathbf{h_{\boldsymbol{\delta}}}^{l-\text{ext}} \mathbf{S}  = {\mathbf{W}}_{\boldsymbol{\delta}} (\mathbf{S}^T \mathbf{\Sigma}_\mathbf{h_{\boldsymbol{\delta}}}^{l-\text{ext}} \mathbf{S} + \sigma_n^2\mathbf{I}_M)
\end{equation}
\begin{equation}
     \mathbf{I}_M  = \mathbf{S}^T {\mathbf{W}}_{\boldsymbol{\delta}} + \sigma_n^2\mathbf{I}_M (\mathbf{S}^T \mathbf{\Sigma}_\mathbf{h_{\boldsymbol{\delta}}}^{l-\text{ext}} \mathbf{S} + \sigma_n^2\mathbf{I}_M)^{-1}
\end{equation}
\begin{equation}
     \mathbf{\Sigma_{y_{\boldsymbol{\delta}}}^{-1}} = \mathbf{\Sigma}_\mathbf{n}^{-1}  (\mathbf{I}_M - \mathbf{S}^T {\mathbf{W}}_{\boldsymbol{\delta}}).
\end{equation}
The likelihood is now re-expressed in terms of ${\mathbf{W}}_{\boldsymbol{\delta}}$ by:
\begin{equation}
  p(\mathbf{y}|{\boldsymbol{\delta}}) \propto \exp{(\sigma_n^{-2}\mathrm{tr}( \mathbf{S}^T {\mathbf{W}}_{\boldsymbol{\delta}} \mathbf{y}\mathbf{y}^H))}|{\mathbf{I}_M - \mathbf{S}^T {\mathbf{W}}_{\boldsymbol{\delta}}|}.
\end{equation}
By defining $\hat{\mathbf{C}} = \sigma_n^{-2}\mathbf{y}\mathbf{y}^H$ and 
\begin{equation}
      b_{\boldsymbol{\delta}}
      = \mathrm{log}|{\mathbf{I}_M - \mathbf{S}^T {\mathbf{W}}_{\boldsymbol{\delta}}|},
      \label{biasTerm}
\end{equation}
the likelihood $p(\mathbf{y}|{\boldsymbol{\delta}})$ is reformulated to:
\begin{equation}
  p(\mathbf{y}|{\boldsymbol{\delta}}) \propto \exp{(\mathrm{tr}( \mathbf{S}^T {\mathbf{W}}_{\boldsymbol{\delta}} \hat{\mathbf{C}})}+b_{\boldsymbol{\delta}}).
\end{equation}
We can now incorporate the result for $p(\mathbf{y}|{\boldsymbol{\delta}})$ into \eqref{estFilt}:
\begin{equation}
\hat{\mathbf{W}}_\text{MMSE} =  \dfrac{\int \exp{(\mathrm{tr}( \mathbf{S}^T {\mathbf{W}}_{\boldsymbol{\delta}} \hat{\mathbf{C}})}+b_{\boldsymbol{\delta}}) {\mathbf{W}}_{\boldsymbol{\delta}} p({\boldsymbol{\delta}})  d{\boldsymbol{\delta}} } { \int \exp{(\mathrm{tr}( \mathbf{S}^T {\mathbf{W}}_{\boldsymbol{\delta}} \hat{\mathbf{C}})}+b_{\boldsymbol{\delta}}) p( {\boldsymbol{\delta}}) d{\boldsymbol{\delta}} }.
\end{equation}
Analogous to \eqref{lstepWPreformu}, the approximated MMSE predictor is: 
\begin{equation}
    \hat{\mathbf{w}}^T(\hat{\mathbf{C}}) 
      =  \mathbf{e}_1^T \dfrac{\int \exp{(\mathrm{tr}( \mathbf{S}^T {\mathbf{W}}_{\boldsymbol{\delta}} \hat{\mathbf{C}})}+b_{\boldsymbol{\delta}}) {\mathbf{W}}_{\boldsymbol{\delta}} p({\boldsymbol{\delta}})  d{\boldsymbol{\delta}} } { \int \exp{(\mathrm{tr}( \mathbf{S}^T {\mathbf{W}}_{\boldsymbol{\delta}} \hat{\mathbf{C}})}+b_{\boldsymbol{\delta}}) p( {\boldsymbol{\delta}}) d{\boldsymbol{\delta}} }.
      \label{filterVector}
\end{equation}
For arbitrary prior distributions $p({\boldsymbol{\delta}})$ this filter cannot be evaluated in closed form \cite{Neumann}. Similar as in \cite{Neumann}, to have a computable expression we make the following assumption \cite{turan2019learning}:\newline
\begin{assumption}
The prior $p({\boldsymbol{\delta}})$ is discrete and uniform: \begin{equation}
    p({\boldsymbol{\delta}_i}) = {1}/{N},  \forall i = 1,\dots N.
\end{equation}
\end{assumption}
We replace the prior in \eqref{filterVector} by ${1}/{N}$ and the integrals by sums and end up with the Gridded Predictor \cite{turan2019learning}:
\begin{equation}
\hat{\mathbf{w}}^T(\hat{\mathbf{C}})  = \mathbf{e}_1^T \dfrac{ ({1}/{N})\sum_{i=1}^{N} \exp{(\mathrm{tr}( \mathbf{S}^T {\mathbf{W}}_{\boldsymbol{\delta}_i} \hat{\mathbf{C}})}+b_{\boldsymbol{\delta}_i}) {\mathbf{W}}_{\boldsymbol{\delta}_i}} { ({1}/{N}) \sum_{i=1}^{N} \exp{(\mathrm{tr}( \mathbf{S}^T {\mathbf{W}}_{\boldsymbol{\delta}_i} \hat{\mathbf{C}})}+b_{\boldsymbol{\delta}_i})},
\label{ass1Filter}
\end{equation}
where the filter of each sample ${\mathbf{W}}_{\boldsymbol{\delta}_i}$ is calculated according to \eqref{lstepWPreformu} and $b_{\boldsymbol{\delta}_i}$ is evaluated by \eqref{biasTerm}.

\subsection{Structured Predictor}
\label{sec:majhead}

With the Gridded Predictor it is possible to achieve prediction without the knowledge of the true PSD of the channel coefficients. The drawbacks are the numerical complexity and a large memory requirement, due to the storage of a filter for each sample ${\mathbf{W}}_{\boldsymbol{\delta}_i}$. By making another assumption, we simplify the predictor and reduce the memory overhead \cite{turan2019learning}:
\begin{assumption}
$\forall i = 1,\dots N$ the filters $\mathbf{S}^T {\mathbf{W}}_{\boldsymbol{\delta}_i}$ can be decomposed as:
\begin{equation}
\mathbf{S}^T {\mathbf{W}}_{\boldsymbol{\delta}_i} = \mathbf{Q}^H \mathrm{diag}(\mathbf{w}_{\boldsymbol{\delta}_i}) \mathbf{Q},
\label{ass2}
\end{equation}
with $ \mathbf{w}_{\boldsymbol{\delta}_i} \in \mathbb{R}^K$ and a common matrix $\mathbf{Q} \in \mathbb{C}^{K \times M}$.
\end{assumption}
It is now sufficient to store a vector $\mathbf{w}_{\boldsymbol{\delta}_i}$ for each sample, which reduces the memory overhead. Possible candidates for $\mathbf{Q}$ are again either $\mathbf{F}_1$ (Circulant approx.) or $\mathbf{F}_2$ (Toeplitz approx.) \cite{Neumann, Gray2006}. By defining $\hat{\mathbf{c}} = {\sigma^{-2}_n} |\mathbf{Qy}|^2$ and using \eqref{ass2}, it follows for the trace expressions in \eqref{ass1Filter}:
\begin{align}
    \mathrm{tr}( \mathbf{S}^T {\mathbf{W}}_{\boldsymbol{\delta}_i} \hat{\mathbf{C}})
      &=  \mathrm{tr}( \mathbf{Q}^H \mathrm{diag}(\mathbf{w}_{\boldsymbol{\delta}_i}) \mathbf{Q} {\sigma^{-2}_n}\mathbf{y}\mathbf{y}^H) \\ 
      &= \mathrm{tr}( \mathrm{diag}(\mathbf{w}_{\boldsymbol{\delta}_i}) {\sigma^{-2}_n}\mathbf{Q} \mathbf{y}\mathbf{y}^H \mathbf{Q}^H ) \\
      &= \mathbf{w}^T_{\boldsymbol{\delta}_i} \hat{\mathbf{c}},
\end{align}
since $\hat{\mathbf{c}}$ contains the diagonal entries of ${\sigma^{-2}_n} \mathbf{Q} \mathbf{y}\mathbf{y}^H \mathbf{Q}^H$. 
The Gridded Predictor from \eqref{ass1Filter} simplifies to:
\begin{equation}
\hat{\mathbf{w}}^T(\hat{\mathbf{c}})  = \dfrac{ \sum_{i=1}^{N} \exp{(\mathbf{w}^T_{\boldsymbol{\delta}_i} \hat{\mathbf{c}}}+b_{\boldsymbol{\delta}_i}) \mathbf{e}_1^T {\mathbf{W}}_{\boldsymbol{\delta}_i}} { \sum_{i=1}^{N} \exp{( \mathbf{w}^T_{\boldsymbol{\delta}_i} \hat{\mathbf{c}}}+b_{\boldsymbol{\delta}_i})}.
\end{equation}
We end up with the \textit{Structured Predictor} of following form \cite{turan2019learning}:
\begin{equation}
\hat{\mathbf{w}}(\hat{\mathbf{c}})  = \mathbf{A}_2 \dfrac{ \exp{(\mathbf{A}_1\hat{\mathbf{c}}+ \mathbf{b} )} } { \mathbf{1}^T\exp{(\mathbf{A}_1\hat{\mathbf{c}}+ \mathbf{b})}},
\label{StrucPred}
\end{equation}
where the matrices $\mathbf{A}_1$ and $\mathbf{A}_2$ and the vector $\mathbf{b}$ are:
\begin{equation}
    \mathbf{A}_1 = 
       \begin{bmatrix}
        \mathbf{w}^T_{\boldsymbol{\delta}_1} \\
        \vdots \\
        \mathbf{w}^T_{\boldsymbol{\delta}_N}
\end{bmatrix} \hspace{0.5cm}
\mathbf{A}_2 = 
\begin{bmatrix}
        \mathbf{e}_1^T\mathbf{W}_{\boldsymbol{\delta}_1} \\
        \vdots \\
        \mathbf{e}_1^T\mathbf{W}_{\boldsymbol{\delta}_N}
\end{bmatrix}^T \hspace{0.5cm}
\mathbf{b} = 
\begin{bmatrix}
        b_{\boldsymbol{\delta}_1}\\
        \vdots \\
        b_{\boldsymbol{\delta}_N}
\end{bmatrix}.
\label{A1A2b}
\end{equation}

\subsection{Neural Network Predictor}
\label{sec:print}
An expert observation shows that a feed-forward neural network with one hidden layer and the softmax activation function, has the same structure as the Structured Predictor. We define the neural network as \cite{turan2019learning}:
\begin{equation}
\hat{\mathbf{w}}_{\text{NN}}(\hat{\mathbf{c}})  = \mathbf{A}_{(2)} \dfrac{ \exp{(\mathbf{A}_{(1)}\hat{\mathbf{c}}+ \mathbf{b}_{(1)} )} } { \mathbf{1}^T\exp{(\mathbf{A}_{(1)}\hat{\mathbf{c}}+ \mathbf{b}_{(1)})}} + \mathbf{b}_{(2)}. 
\end{equation}
The matrix $\mathbf{A}_1$ of the Structured Predictor from \eqref{StrucPred} is equal to the weight matrix $\mathbf{A}_{(1)}$ of the first layer: $\mathbf{A}_{(1)} = \mathbf{A}_1$.
The vector $\mathbf{b}$ is the bias vector of the first layer, thus $\mathbf{b}_{(1)} = \mathbf{b}$.
$\mathbf{A}_2$ consists of sample specific filter vectors $\mathbf{e}_1^T{\mathbf{W}}_{\boldsymbol{\delta}_i} \in \mathbb{C}^{1\times M}$. Thus, the matrix $\mathbf{A}_2$ is complex. Therefore, we split the matrix $\mathbf{A}_2$ into its real and imaginary part and define 
\begin{equation}
     \mathbf{A}_{(2)} = 
     \begin{bmatrix}
        \Re(\mathbf{A}_2) \\
        \Im(\mathbf{A}_2) \\
\end{bmatrix}.
\label{NN3}
\end{equation}
We further define a bias term for the second layer and the Structured Predictor suggests: $\mathbf{b}_{(2)} = \mathbf{0}$. Accordingly, the output of the neural network $\hat{\mathbf{w}}_{\text{NN}}(\hat{\mathbf{c}})$ is the concatenation of the real and imaginary parts of the Structured Predictor $\hat{\mathbf{w}}(\hat{\mathbf{c}})$:
\begin{equation}
    \hat{\mathbf{w}}_{\text{NN}}(\hat{\mathbf{c}}) = 
         \begin{bmatrix}
        \hat{\mathbf{w}}_{\text{NN},\Re}(\hat{\mathbf{c}}) \\
        \hat{\mathbf{w}}_{\text{NN},\Im}(\hat{\mathbf{c}}) \\
\end{bmatrix} = 
\begin{bmatrix}
        \Re(\hat{\mathbf{w}}(\hat{\mathbf{c}})) \\
        \Im(\hat{\mathbf{w}}(\hat{\mathbf{c}})) \\
\end{bmatrix}.
\end{equation}
Thus, we obtain an estimate (of a specific channel realization, indexed by $b$) by calculating: 
\begin{equation}
\hat{h}_{b,\mathcal{I}_N} = [\hat{\mathbf{w}}_{\text{NN},\Re}(\hat{\mathbf{c}}_b) + j\hat{\mathbf{w}}_{\text{NN},\Im}(\hat{\mathbf{c}}_b)]^T\mathbf{y}_{b,\mathcal{I}_M}.    
\end{equation}
The learning procedure is described in the following \cite{turan2019learning}:
\begin{algorithm}[H]
\caption{Learning the MMSE Channel Predictor}
\begin{algorithmic}[1]
\STATE Init. the Neural Network with the Structured Predictor
\STATE Generate a mini-batch of in total $B$ channel realizations, of the observation interval $\mathbf{h}_{b,\mathcal{I}_M}$ and corresponding channel coefficients of the prediction interval (of desired prediction step $l$) $h_{b,\mathcal{I}_N}$, for $b = 1,2,\dots,B$.
\STATE Generate noisy version $\mathbf{y}_{b,\mathcal{I}_M}$ of $\mathbf{h}_{b,\mathcal{I}_M}$ and calculate $\mathbf{\hat{c}}_b$ (input of the neural network), for $b = 1,2,\dots,B$.
\STATE Calculate stochastic gradient ($i = 1,2$):
\begin{equation*}
    \mathbf{g} = \dfrac{1}{B} \sum_{b=1}^{B} \dfrac{\partial}{\partial[\mathbf{A}_{(i)};\mathbf{b}_{(i)}] } \| h_{b,\mathcal{I}_N} -  \hat{h}_{b,\mathcal{I}_N}\|_2^2,
\end{equation*}
\STATE Update the variables of the neural network with a desired gradient algorithm (e.g., \cite{Kingma})
\STATE Repeat steps 2-5 until a convergence criterion is fulfilled.
\end{algorithmic}
\end{algorithm}

\subsection{Neural Network based Channel Prediction in an Indoor Scenario}
\label{sec:illust}
In the following, we use the ray-tracing scenario "I1" of the DeepMIMO dataset with a carrier frequency of $2.4\si{GHz}$ at BS8 with one antenna. We consider one carrier and at all possible grid points; thus, the first active row is R1 and the last active row is R502. In this scenario the horizontal and vertical distances between neighbouring grid points are $1\si{cm}$. The channel coefficients are normalized such that $E[\| \mathbf{h} \|^2] = 1$. The obtained channel vectors of each grid point are grouped together, to emulate a moving user ($v = 4\si{km/h}$) along the positive x-axis direction, as follows: Five consecutive channel coefficients are stacked together, where the first four form the observation interval channel vector $\mathbf{h}$ and we want to predict the fifth channel coefficient; thus, $M=4$ and $l=1$. We randomly assigned the groups of channel coefficients either to a training set or a test set. Thus, we have 500 training batches, where each consists of 50 channel realizations and 103 test batches, with each including 50 channel realizations. We train over 20 epochs by shuffling the train dataset each time. 
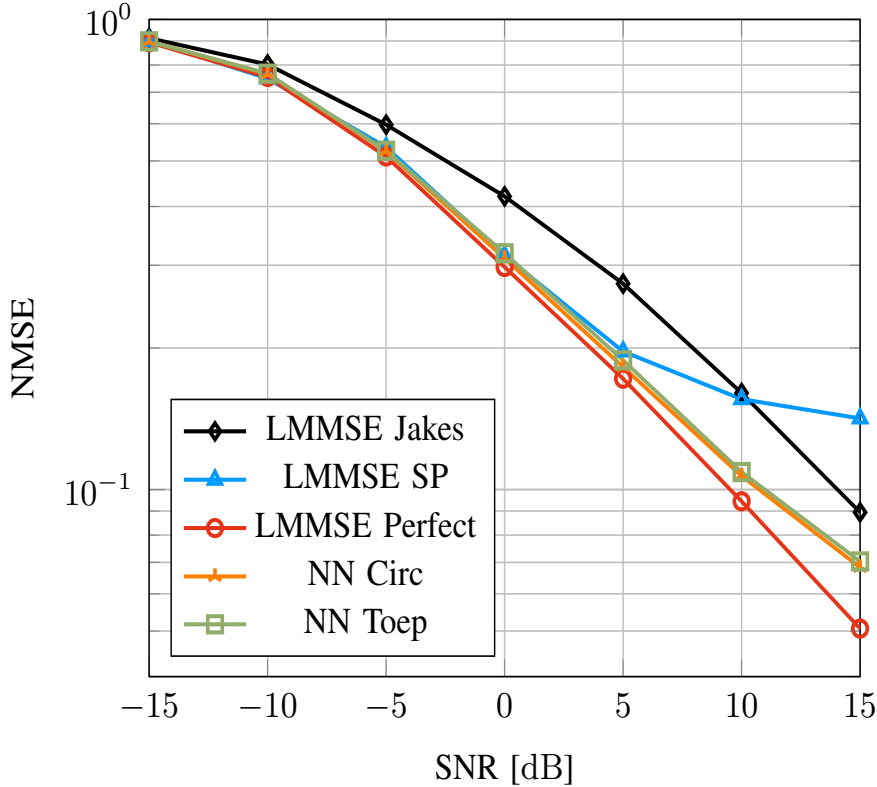
\begin{figure}[t]
    \centering
    \resizebox{350pt}{308pt}{
    \begin{tikzpicture}
\begin{axis}[
legend pos= south west,
legend style={font=\footnotesize},
xmin=-15,
xmax=15,
ymin=4e-2,
ymax=1e+0,
xlabel={SNR [\si{dB}]},
ylabel={NMSE},
grid=both,
ymode=log
]
\addplot[line width=1,solid,mark=diamond]
table[x=snr,y=nmse] {LCP_25_LMMSEJakes.txt};
\addlegendentry{LMMSE Jakes};
\addplot[line width=1,TUMBeamerBlue,solid,mark=triangle]
table[x=snr,y=nmse] {LCP_25_LMMSE1P.txt};
\addlegendentry{LMMSE SP};
\addplot[line width=1,TUMBeamerRed,solid,mark=o]
table[x=snr,y=nmse] {LCP_25_LMMSEPerfect.txt};
\addlegendentry{LMMSE Perfect};
\addplot[line width=1,TUMBeamerOrange,solid,mark=Mercedes star]
table[x=snr,y=nmse] {LCP_25_NNCirc.txt};
\addlegendentry{NN Circ};
\addplot[line width=1,TUMBeamerGreen,solid,mark=square]
table[x=snr,y=nmse] {LCP_25_NNToep.txt};
\addlegendentry{NN Toep};
\end{axis}
\end{tikzpicture}
}
\caption{NMSE at prediction step l = 1 with M = 4 at BS8.}
\label{fig:LCP25}
\end{figure}

In the simulation of Fig. \ref{fig:LCP25} the number of paths is set to the maximum value of 25 (all available paths are used to construct the channel coefficient of each grid point). Over the simulated observation length of $M=4$ and the additional prediction coefficient ($l=1$) the DoAs fluctuate by approximately $\pm 2 \degree$. Thus, we use as baseline the LMMSE predictor with (almost) perfect knowledge of the statistical moments of second order based on the covariance function $ R_h[k]$ (\textit{LMMSE Perfect}). We further use as benchmark the LMMSE perdictor, where we construct the covariance matrix by only considering the strongest path, and denote it by \textit{LMMSE SP}. The LMMSE predictor with the assumption of $P \to \infty$, is denoted as \textit{LMMSE Jakes}. The Neural Network Predictor is denoted as \textit{NN Toep} or as \textit{NN Circ}, where we initialize the weight matrices and biases with the corresponding Structured Predictors, which are constructed by using the channel model from \eqref{channelcoeff} with only one propagation path $P=1$ and with $N=4$ (Circulant approx.) or $N=8$ (Toeplitz apporx.), and then train with the data provided by the indoor scenario of the DeepMIMO dataset. We have constructed the Structured Predictors with $P=1$, since we assume to have one dominating path, which is the path with a line of sight to the BS. The predictors are specifically constructed for each simulated SNR, i.e., we have to construct and train the neural network predictors for each SNR separately.

The neural network predictors trained with the indoor scenario data, do not only compensate the approximation errors of Assumption 1 and Assumption 2, instead it is possible to predict the channel for a large SNR range clearly better than compared to the LMMSE Jakes predictor (Fig. \ref{fig:LCP25}). We can see further that, the NN Toep and NN Circ predictor reach a similar NMSE as compared to the LMMSE SP predictor for the SNR range from $-15\si{dB}$ to $0\si{dB}$ (Fig. \ref{fig:LCP25}) and outperform the LMMSE SP predictor for all higher SNR values. The LMMSE SP predictor performs bad for high SNR values, because in the construction of this predictor the strongest path is considered exclusively. However, for higher SNR values the sub-path powers are not negligible as compared to the noise and need to be incorporated and reflected within the channel coavariance matrix accordingly. The LMMSE Perfect predictor is constructed by considering all paths and with the neural network predictors we can achieve a similar performance, especially for low SNR values (Fig. \ref{fig:LCP25}).

\section{Conclusion}
We evaluated the performances of the convolutional neural network based channel estimators from \cite{Neumann} in a mmWave environment using data provided by the DeepMIMO dataset and confirmed, that these estimators work in a ray tracing based simulation environment in addition to the 3GPP based model and real measurement data as in \cite{Hellings}. 

A novel approach to learn a feed-forward neural network channel predictor was further presented in this paper. Starting from the LMMSE predictor a reformulated version was derived. By making two key assumption it was possible to further derive predictors, which are motivated by the structure of the MMSE predictor. The neural network predictor is initialized with the Structured Predictor. By further training the network based predictors, it is possible to compensate the approximation errors due to the assumptions we made. Simulation results show that the neural network predictors outperform the LMMSE predictor based on the assumption of a Jakes spectrum and achieve a similar NMSE as compared to the LMMSE predictor with perfect knowledge of the statistical moments of second order based on the covariance function, especially for low SNR values.

We further guaranteed the reproducibility of our work by using the data provided by the generic DeepMIMO dataset in our simulations. 
\vfill

\bibliographystyle{IEEEtran}
\bibliography{Turan_Utschick_Reproducible_NN_Evaluation}

\end{document}